\documentclass[aps,prl,twocolumn,floatfix,showpacs]{revtex4}

\newcommand{\beq}[1]{\begin{eqnarray} \label{#1}}
\newcommand{\eeq}{\end{eqnarray}}

\newcommand{\etal}{\emph{et al.}, }
\newcommand{\ibid}[1]{\emph{ibid.} \textbf{#1}, }
\newcommand{\prlj}[1]{\prl{\textbf{#1}},}
\newcommand{\prbj}[1]{\prb{\textbf{#1}},}

\newcommand{\PGRPH}[1]{}

 \newcommand{\SHOWFIG}[1]{\epsfig{figure=#1, width=1\linewidth,clip=}}

\newcommand{\COMMENTJ}[1]{}

\usepackage{amssymb}
\usepackage{epsfig}
\usepackage{amsmath}

\begin{document}

\title{Spin-texture and magneto-roton excitations at $\nu=1/3$  }  

\author{Javier G. \surname{Groshaus} $^{1,2}$}
\email{jgg@phys.columbia.edu}
\altaffiliation{Institute for Optical
Sciences, Dept. of Chemistry \& Dept. of Physics, University of
Toronto, Toronto, ON, Canada.}
\author{Irene \surname{Dujovne}$^{1,2}$}
\altaffiliation{TU Delft, Kavli Inst. of Nanoscience, The
Netherlands.}
\author{Yann \surname{Gallais}$^{1}$}
\altaffiliation{Laboratoire Mat\'eriaux et Ph\'enom\`enes
Quantiques, CNRS UMR 7162, Universit\'e Paris 7, France.}
\author{Cyrus F. \surname{Hirjibehedin}$^{1,2}$}
\altaffiliation{London Centre for Nanotechnology, Depts. of Physics
\& Astronomy and Chemistry, UCL, London, U.K.}
\author{Aron \surname{Pinczuk} $^{1,2}$}
\author{Yan-Wen \surname{Tan}$^{1}$}
\altaffiliation{Dept. of Physics, UC Berkeley, CA 94720.}.
\author{Horst \surname{Stormer}$^{1}$}
\affiliation{$^1$ Physics \& Applied Physics and Applied
Mathematics, Columbia University, New York, NY 10027}
\author{Brian S. \surname{Dennis} $^2$}
\author{Loren N. \surname{Pfeiffer} $^2$}
\author{Ken  W. \surname{West} $^2$}
\affiliation{$^2$ Alcatel-Lucent Bell Labs, Murray Hill, NJ 07974}

\date{\today}

\begin{abstract}
Neutral spin texture ST excitations at $\nu=1/3$ are directly
observed for the first time by resonant inelastic light scattering.
They are determined to involve two simultaneous spin-flips. At low
magnetic fields, the ST energy is below that of the magneto-roton
minimum. With increasing in-plane magnetic field these modes
energies cross at a critical ratio of the Zeeman and Coulomb
energies of $\eta_c=0.020 \pm 0.001$. Surprisingly, the intensity of
the ST mode grows with temperature in the range in which the
magneto-roton modes collapse. The temperature dependence is
interpreted in terms of a competition between coexisting phases
supporting different excitations. We consider the role of the ST
excitations in activated transport at $\nu=1/3$.
\end{abstract}

\pacs{73.43.Lp, 73.43.Nq, 73.43.-f, 73.20.Mf}

\maketitle

\PGRPH{1 - intro.1}

Collective excitations of two dimensional electron systems (2DES)
are a revealing probe into the physics of fractional quantum Hall
(FQH) fluids. A common approach to probe collective modes in the
FQH regime consists of measuring the thermal activation energy of
the longitudinal resistivity, $\Delta_a$, as done in Refs.
\cite{Boebinger85, Leadley97, Kukushkin00, DethlefsenWojs2006}. At
filling factor $\nu=1/3$, the transport activation mechanism is
often attributed to neutral excitations in the charge degree of
freedom, namely, magneto-roton modes
 in the
limit of large wavevector $\Delta_{\infty}$ \cite{Haldane85,
Pinczuk93, Turberfield97, Scarola2000, Morf2002}. The magneto-roton
wavevector dispersion at $\nu=1/3$ is depicted in Fig. 1a. The
magneto-roton energies scale with the Coulomb energy
$E_C=e^2/\epsilon l_B$, where $\epsilon$ is the dielectric constant,
$l_B = (h c/2 \pi eB_{\perp})^{1/2}$ is the magnetic length and
$B_{\perp}$ is the magnetic field normal to the 2DES. Thus $E_C
\propto \sqrt{B_{\perp}}$.

\PGRPH{2 - intro.2}

Even in the cleanest samples, the ones with minimal residual
disorder, the measured activation gap is significantly smaller
than the calculated energies for $\Delta_{\infty}$
\cite{Scarola2000, Morf2002}. Moreover, for low values of the
ratio of the Zeeman and Coulomb energies, $\eta=E_z/E_c$, at
$\nu=1/3$ the measured $\Delta_a$ fails to scale with
$\sqrt{B_{\perp}}$. $\Delta_a$ presents a term increasing linearly
with the Zeeman energy, $E_z = g \mu B$, where $B$ is the total
magnetic field. More surprisingly, $\Delta_a$ grows with $E_z$
with a slope $s \equiv
\partial \Delta_a /
\partial E_z=3$ \cite{Leadley97} or $s=2$ \cite{Kukushkin00,
DethlefsenWojs2006}. Such energy evolution suggests that at low
$\eta$, the relevant thermally excited modes are spin-texture
modes (ST), namely, modes involving $s$ simultaneous spin-flips.
Upon excitation of an ST mode, the component along the magnetic
field of the the total spin of the 2DES $S_z$ is decreased by
$\Delta S_z=\hbar s$.

\begin{figure}[h!]
\SHOWFIG{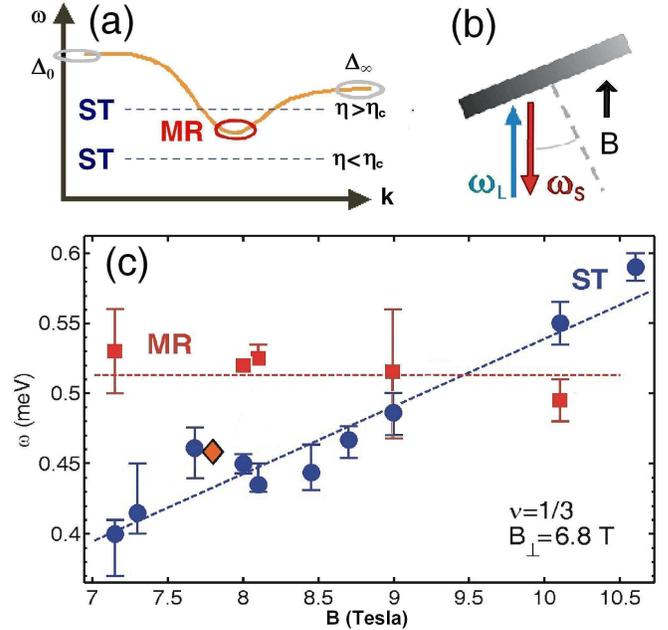}
 \caption{(color online)
    (a) Dispersion of the lowest magneto-roton branch for excitation
in the charge degree of freedom at $\nu=1/3$. The ST is lower in
energy than the MR for $\eta<\eta_c$ and larger at $\eta>\eta_c$.
(b) Experimental setup. (c) Energies of the ST and MR as the total
field $B$ increases at constant $B_{\perp}$. Circles (squares)
denote peaks resonant with $S_1$ ($T_B$). The diamond is the
activated transport gap for $B=B_{\perp}=7.8$ T at $\nu=1/3$. Dashed
lines are fits. } \label{fig:1}
\end{figure}

 \PGRPH{3 - intro.3}

Transport measurements at $\nu=1$ for varying values of the
in-plane component of the field $B_{//}$ unveiled a steep change
of $\Delta_a$ with $E_z$, with a slope $s$ much larger than at
$\nu=1/3$. The activation mechanism was attributed to ST modes
consisting of a Skyrmion-Antiskyrmion neutral pair created at
$\nu=1$. These modes involve the flipping of $s$ spins
\cite{Schmeller95, Nicholas98}. Skyrmion and Antiskyrmions were
predicted to be the ground state of the system at low $\eta$ and
away from $\nu=1$ by one magnetic flux quantum. Such states are
described as a radial distribution of spin density and charge,
with a total charge of $-|e|$ for $\nu>1$ (Skyrmion), and $+|e|$
for $\nu>1$ (Antiskyrmion) \cite{Sondhi93, Fertig94}. Spin
textures
 have been invoked also to explain the ground state around
$\nu=1/3$ and the activation energies at $\nu=1/3$ \cite{Sondhi93,
Kamilla96, Ahn97, Mandal98, Doretto2005, DethlefsenWojs2006}.

\PGRPH{4 - intro.4}

Resonant inelastic light scattering is a direct probe of
collective excitations of FQH liquids. Light scattering
experiments have identified the long wavelength magneto-roton mode
$\Delta_0$ \cite{Pinczuk93,Turberfield97}, modes at the
magneto-roton minimum (MR) and large wave-vector modes.
\cite{Kang2001,CyrusPRL2003,Dujovne2003}. The experimental picture
arising from these light scattering measurements is still far from
complete. Several peaks were observed near the MR energy and their
identification is not unequivocal \cite{CyrusPRL2003}. Moreover,
no ST excitations have yet been found in inelastic light
scattering either at $\nu=1$ nor at $\nu=1/3$. Measurements of the
spin polarization were not able to provide conclusive evidence for
STs in the ground state away from $\nu=1/3$ \cite{Khandelwal98}.

\PGRPH{5 - thisletter.1}

We report here the first direct observation of a neutral
spin-texture by resonant inelastic light scattering at $\nu=1/3$.
Identification of the ST mode comes from the observation that as we
raise $B_{//}$ (and thus $\eta$) at $\nu=1/3$, the energy of the ST
increases and crosses that of the MR at a field of $\sim 9.5$ T
corresponding to $\eta_c = 0.020 \pm 0.001$, as shown in Fig. 1c.
The dependence of the ST energy $\omega_{ST}$ on $B$ indicates that
the ST mode involves $s \equiv
\partial \omega_{ST} / \partial E_z = 2$ spin flips.

\PGRPH{6 - thisletter.2}

A surprising property of the discovered ST mode is that its
intensity is greatly enhanced with increasing temperature. This
behavior contrasts that of the magneto-roton modes, which collapse
in the same temperature range ($\sim 0.2-1$ K). These results are
consistent with the coexistence of phases supporting magneto-roton
excitations and phases supporting ST modes.

\PGRPH{5 - thisletter.2}

\PGRPH{7 - technique.1}

Optical measurements were performed on a single side modulation
doped Al$_{.06}$Ga$_{.94}$As/GaAs 33 nm quantum well with electron
density $n=5.3\times10^{10}$ cm$^{-2}$  and mobility
$\mu=7.2\times10^{6}$ cm$^{2}/$Vs. The sample was mounted on the
cold finger of a dilution fridge with optical windows and cold
finger temperatures reaching 40 mK. The sample was mounted at an
angle with respect to the magnetic field $B$ (see Fig. 1b). The
value of $\eta$ is changed by varying this angle and adjusting $B$
to keep $\nu=1/3$. This procedure amounts to varying $B_{//}$ at
constant $B_{\perp}$. A laser beam of photon energy $\omega_L$ is
incident on the sample along $B$, and the backscattered photon, of
energy $\omega_S$, is dispersed in a double spectrometer and
recorded with a CCD camera. By energy conservation, the energy of
the excited mode in the 2DES is given by the energy shift $\omega =
\omega_L - \omega_S$. The in-plane wave-vector is not strictly
conserved due to the presence of weak disorder. Hence, this
technique allows for the excitation of relatively large density of
states modes of wave-vector larger than that provided by photon
recoil, such as the MR (see Fig. 1a). Transport measurements of
$\Delta_a$ were performed on a similar Al$_{0.1}$Ga$_{0.9}$As/GaAs
33 nm quantum well with $n=6.3\times10^{10}$ cm$^{-2}$ and
$\mu=14\times10^{6}$ cm$^{2}/$Vs in a perpendicular magnetic field.

\PGRPH{8 - Explain Fig. 2ab}

Figs. 2a and 2b show resonant light scattering spectra at
$\nu=1/3$ and total field $B=8$ T which is in the low $\eta$
region ($\eta<\eta_c$). It can be seen that although the MR and ST
modes are very close in energy, the modes can be selectively
excited by varying $\omega_L$. This selectivity is due to
intermediate virtual states in the scattering process that are
resonant with the scattered photon energy $\omega_S$. A similar
outgoing resonance has been observed for the long wavelength
spin-wave \cite{CyrusResonance2003}.

\PGRPH{8 - Explain Fig. 2cd}

The outgoing resonance is illustrated in Fig. 2c, where we present a
compilation of spectra such as those in Fig. 1(a,b) as a function of
$\omega$ and $\omega_S$. The intensity of the signal is color coded
(dark red means high intensity). Each mode, ST or MR, is identified
by its corresponding $\omega$. In addition, it can be seen that the
ST and MR modes become resonant at different values of $\omega_S$.
Figure 2d shows the profiles of resonant enhancement of the
intensities of the ST (blue dots) and MR modes (red dots) and the
photoluminescence spectrum (PL, black line). The PL is mainly
composed of two peaks, $S_1$ and $T_B$, which have been extensively
studied. They are associated with singlet and triplet excitonic
states, respectively \cite{Yusa2001, CyrusResonance2003}. Figures 2c
and 2d reveal that the profile of resonant enhancement of the ST
mode overlaps with $S_1$ while that of the MR peaks at the $T_B$
energy.

\PGRPH{9 - Explain Fig. 3}

Selective excitation of the ST and MR modes enables us to follow
the evolution of their energy as $B_{//}$ is increased. Fig.
 3 shows measurements at a higher $B_{//}$ corresponding to
 $\eta>\eta_c$.
 These measurements reveal that while the energy of the
  MR mode has hardly changed, the ST has increased
 in energy above that of the MR.

\begin{figure}
\SHOWFIG{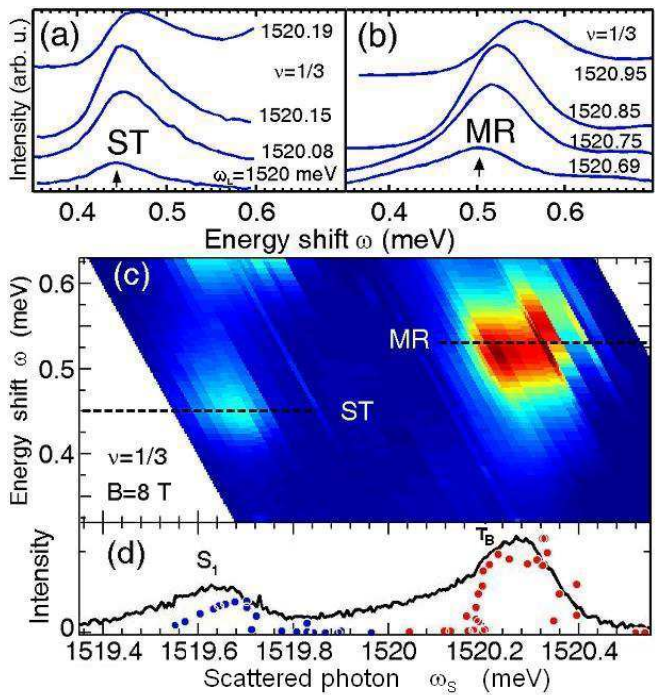}
 \caption{ (color) (a,b) Excitation spectra for
different laser energies $\omega_L$ at 70 mK, $\nu=1/3$ and total
field $B=8$ T. (c) Spectra such as those in (a,b) as a function of
$\omega$ and $\omega_S$. Dark red means high intensity. (d)
Resonant enhancement profiles of the ST and MR modes (dots)
superimposed to the photoluminescence spectrum (black line). }
\label{fig:2}
\end{figure}

\begin{figure}
\SHOWFIG{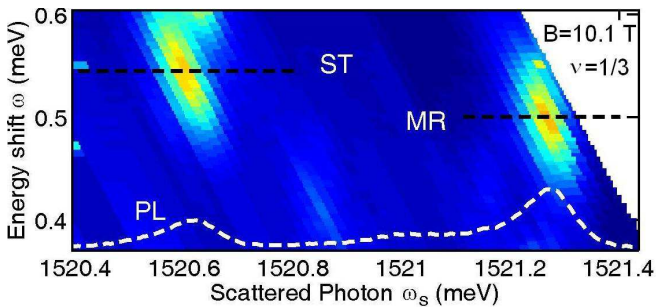} \caption{ (color online) Spectra at 40 mK
 and $\nu=1/3$ as in Fig. 2c, but at the higher total field of $B=10.1$ T
($B_{\perp}$ is unchanged). The white line is the
photoluminescence spectrum (PL).} \label{fig:3}
\end{figure}

\PGRPH{10 - Explain Fig 4 -  Temperature dependence}

Figure 4 shows that the ST and magneto-roton modes have strikingly
different temperature behaviors.  As the temperature is raised, the
strength of the ST mode increases while that of the MR and
$\Delta_0$ decreases. The quenching of $\Delta_0$ with increasing
temperature has been observed before \cite{Pinczuk93}. A higher
energy mode at $0.75$ meV at $B=7.15$ T, possibly corresponding to
 $\Delta_{\infty}$, shows a similar quenching behavior as the
MR and $\Delta_0$ (not shown). No change in energy or in width is
observed for the peaks associated with any of the aforementioned
modes.

\PGRPH{11 - Explain Fig 4 -  Helium analogy, conjecture}

The quenching of the magneto-rotons at $\nu=1/3$ is reminiscent of
the observed quenching of the rotons in superfluid Helium excitation
spectra \cite{Greytak69}. In those experiments, the strength of the
peak was fitted by a line proportional to $n_0(T) = n_0(0)[1 -
(T/T_c)^{\alpha}]\Theta(T_c-T)$, where $n_0(T)$ was interpreted to
be the fraction of particles that have condensed at a given
temperature, $T_c$ and $\alpha$ are fitting parameters and
$\Theta(x)$ is the Heaviside step function. Helium analogies are
supported by a formulation of the theory that maps the 2DES to a
system of interacting bosons in the ground state at $\nu=1/3$
\cite{Girvin87}. Fig. 4 shows that the intensity of the MR and
$\Delta_0$ can be well fitted as proportional to $n_0(T)$, with a
quenching occurring near $T_c=0.85$ K. Remarkably, the ST intensity
starts saturating at the same characteristic temperature at which
the magneto-rotons quench. Therefore, we fit the ST with a dashed
line proportional to $1-n_0(T)$ using the same $\alpha$ and $T_c$
that we used for the magneto-rotons. It is noteworthy that the
temperature range in which the magneto-roton modes survive largely
overlaps the limited range at which longitudinal resistivity
presents activated behavior \cite{Boebinger85, Kukushkin00,
DethlefsenWojs2006}. These results are consistent with a weakly
disordered inhomogenous quantum fluid \cite{Efros1992} in which
phases supporting magneto-roton excitations occupy a fraction
$n_0(T)$ of the sample and coexist with and phases supporting ST
modes in the remaining area.

\PGRPH{10 - exp.4}

\begin{figure}
\SHOWFIG{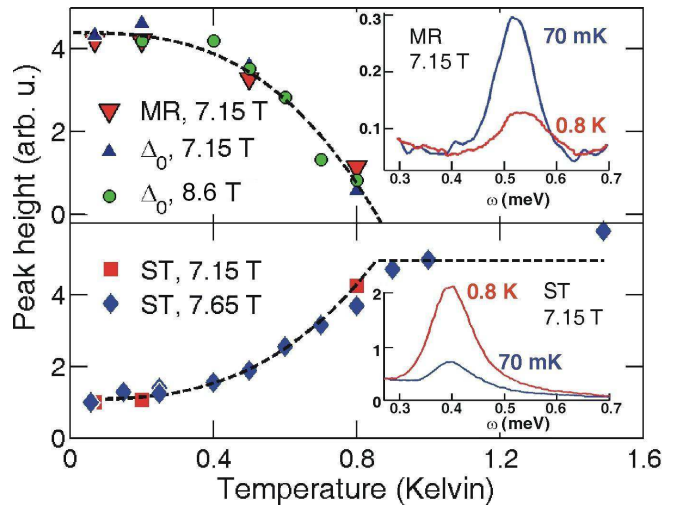}
 \caption{
(color online) Temperature dependence of the intensity of the MR,
$\Delta_0$ and ST peaks at $\nu=1/3$ at different values of $B$.
Each mode was normalized independently for clarity. The dashed lines
are fits with $T_c=0.85$ K, $\alpha=2.84$, $n_0(0)=0.78$. (Insets)
Spectra at two temperatures. } \label{fig:4}
\end{figure}

\PGRPH{12 - discuss.1 - filling factor dependence}

We also find that the MR and $\Delta_0$ modes are rapidly quenched
as the filling factor is tuned away from $1/3$. Such behavior has
been reported before for the $\Delta_0$ \cite{Pinczuk93}. These
findings suggest that as the system is rendered compressible, the
quantum fluid that supports magneto-roton excitations disappears. We
note that a mode with energy close to the MR has been reported to
exist for a broad range of filling factors down to $\nu=1/5$
\cite{Kang2001, CyrusPRL2003}. We find this mode to be resonant with
$S_1$. At $\nu=1/3$, it corresponds to the ST mode.

\PGRPH{13 - fittings in Fig. 1}

Turning again to Fig. 1c, we find that the MR mode energy remains
approximately constant at $0.046 E_C$. This value is close to the
calculated value of $0.052 E_C$ including finite width effects
\cite{Scarola2000,DethlefsenWojs2006}, and consistent with the
experimental value of $0.045 E_C$ obtained from $\Delta_a$
measurements for $\eta>\eta_c$ \cite{DethlefsenWojs2006}. Disorder
may cause a slight decrease in the MR energy as $B$ increases
\cite{Murthy}.

Our activated transport measurements at $\nu=1/3$ in a similar
sample yield $\Delta_a=5.3$ K, as shown in Fig. 1c (diamond). It is
seen that the value $\Delta_a$ is consistent with the energy of the
ST mode (both $\Delta_a$ and $E_z$ are expressed in units of the
corresponding $E_C$). The ST energy is also consistent (within
$<15$\%) with the latest published results for $\Delta_a$ in the
same range of $\eta$, for $\eta<\eta_c$ \cite{Leadley97,
Kukushkin00, DethlefsenWojs2006}.

The energy of the ST mode in Fig. 1c is well fitted by the linear
expression $\omega_{ST} = s E_z + E_0$ with $s=2$ and $E_0=0.06$
meV$=0.005E_c$. We stress that for an accurate extraction of $s$,
the value of $E_z$ as measured \emph{in situ} by inelastic light
scattering was used \cite{Pinczuk93, Turberfield97,
 CyrusResonance2003}, resulting in a $g$-factor value of $g=0.41$,
 lower than the bulk value of $g=0.44$. Experimentally, $E_0$
 is given by the extrapolated $E_z=0$ intercept. This value of
  $E_0$ is consistent with the values obtained by linearly extrapolating
  to $E_z=0$ the published values of $\Delta_a$ at $\nu=1/3$ for Ref. \cite{Kukushkin00}
(with $s=2$), and Refs. \cite{Leadley97, Nicholas98} (with $s=3$).
In Ref. \cite{DethlefsenWojs2006}, $\Delta_a$ extrapolates at
$E_z=0$ to a negative value.

\PGRPH{14 - discuss.2 - contradiction with theoretical}

The experimental values of $E_0$ are significantly lower than
theoretical estimates of $E_0$ for Skyrmions and Antiskyrmions at
$\nu=1/3$. In Ref. \cite{DethlefsenWojs2006}, the activation of the
resistivity process at low $\eta$ was described as the creation of a
spin-reversed quasiparticle and a small Antiskyrmion in which one
additional spin is flipped \cite{DethlefsenWojs2006}, yielding a
total $s=2$. The flipping of an extra spin to form an Antiskyrmion
has a cost of $E_z$ in energy, while simultaneously obtaining a
Coulomb gain in energy. $E_0$ can be obtained as the difference
between the Coulomb gain that results from flipping an extra spin
and the energy cost (due to loss of exchange energy) to spatially
separate a quasihole from a spin-reversed quasiparticle
$\Delta^{\uparrow \downarrow}$ ($\Delta^{\uparrow \downarrow} \sim
0.05 E_C$ \cite{Dujovne2003}). Calculations have been done both the
strict 2D limit and including finite width effect
\cite{DethlefsenWojs2006, Kamilla96, Rezayi87, Mandal2001, Morf2002,
Doretto2005, Sondhi93}. For $s=2$, $E_0$ is predicted to be in the
range $0.024-0.061 E_c$, much higher than observed.

A similar discrepancy with theory is found in the case of $\nu=1$,
where the Coulomb energy is expected to be large for a
Skyrmion-Antiskyrmion pair, but nevertheless $E_0$ varies widely
among samples, and even vanishes \cite{Schmeller95, Nicholas98}. The
low value of $E_0$ is often accounted for by a negative term
attributed to the existence of disorder that lowers the gap
\cite{DethlefsenWojs2006}. The role of disorder in transport and in
optical measurements is not well understood and is still a subject
of active research \cite{Yacoby2004}. If indeed the energy of the ST
is lowered due to disorder, such effect would be stronger at
incompressible fractions where the disorder is not screened. This is
consistent with our observation that the energy of the ST mode
presents minima at $\nu=1/3$ and $2/7$ (not shown).

\PGRPH{conclusion}

In conclusion, we have identified an ST mode in which two spins
are flipped upon inelastic light scattering at $\nu=1/3$. We found
a crossover between the ST mode and the MR mode as $B$ is
increased. The ST and the magneto-roton peaks present strikingly
opposite temperature behavior. The ST mode energy is consistent
with the activation energies for activated transport.

\PGRPH{ACKNOWLEDGEMENT}

 This work was supported by the National Science
Foundation (NSF) Grant No. DMR-03-52738, the Department of Energy
Grant No. DE-AIO2-04ER46133, the Nanoscale Science and Engineering
Initiative of the NSF Grants No. CHE- 0117752 and No. CHE-0641523,
the New York State Office of Science, Technology, and Academic
Research, and the W. M. Keck Foundation.


\end{document}